\documentclass[letterpaper,11pt,leqno]{article}
\usepackage{paper}
\hypersetup{pdftitle={Night and Day: Diurnal Warming and Structural Transformation in India}}

\newcommand{\bib}{references.bib}

\newcommand{\tmin}{\ensuremath{T_{\mathrm{min}}}}
\newcommand{\tmax}{\ensuremath{T_{\mathrm{max}}}}
\newcommand{\btmin}{\ensuremath{\beta_{T_{\mathrm{min}}}}}
\newcommand{\btmax}{\ensuremath{\beta_{T_{\mathrm{max}}}}}

\newcommand{\bdtmin}{\ensuremath{\beta_{\Delta T_{\mathrm{min}}}}}
\newcommand{\bdtmax}{\ensuremath{\beta_{\Delta T_{\mathrm{max}}}}}

\graphicspath{{./Figures/}}

\providecommand{\degree}{\ensuremath{^{\circ}}}

\begin{document}

\title{Night and Day: Diurnal Warming and\\ Structural Change}

\author{Vedarshi Shastry \\ University of California, Santa Cruz
    \thanks{
    Working paper; latest version at
    \href{https://arxiv.org/abs/2607.00279}{[arXiv:2607.00279]}.
    This research was conducted and written independently with no 
    direct financial support or conflict of interest.
    All errors are my own.
    I am grateful to Aaditya Dar and Ajay Shenoy for their feedback
    over various stages of this project, alongside seminar participants
    and discussants at the World Bank, UC Santa Cruz and Royal Economic Society.
    This paper has also significantly benefited from AI agentic research
    assistance.
    \\
    Email: \texttt{veshastr@ucsc.edu}
    }
}

\date{\today}

\begin{titlepage}
\maketitle

\begin{abstract}
This paper shows how the composition of local economic activity 
  shifts decadally under diurnal warming---higher nighttime and daytime temperatures---
  within Indian districts (1981--2011). 
Warmer nights raise, warmer days lower 
  agricultural wage-labour shares.
Though warmer days and nights both contract grain 
  production and area under cultivation,
only warmer days raise local harvest prices.
Long differences show the labour divergence 
  is within the rural population.
In towns, both margins depress non-agricultural worker shares.
These results are consistent
  with a model where warmer days shock land and labour productivity, but
  warmer nights are land-biased shocks in agriculture.
\end{abstract}

\vspace{1.5em}
\noindent \textbf{Keywords:} Climate Change, Structural Transformation, Agriculture, India \\
\noindent \textbf{JEL Codes:} O1, Q54, J43, R11

\end{titlepage}


Structural transformation---the reallocation of labour from agriculture into non-agricultural
employment---is widely documented as the foundation of economic growth and development 
\citep{Lewis1954, Matsuyama1992Dec, Kongsamut2001Oct, Herrendorf2014}.
A growing literature studies how climate change impacts structural transformation
in developing economies \citep{Saxena2024, DasguptaRaghav2024, Pham2026}.
In particular, rising temperatures have been shown to have greater impacts 
than rainfall fluctuations on agricultural productivity \citep{Schlenker2009, Dell2014} 
and labour reallocation \citep{Colmer2021, LST2023}.
Yet, the evidence diverges on whether warming temperatures raise or lower
agricultural labour shares, which has significant implications on structural change.

On the one hand, 
\citet{Jessoe2018} find extreme heat in rural Mexico depresses formal wage employment 
and pushes out-migration.
In India, though heat shocks increase temporary out-migration \citep{Das2026},
caste-based insurance networks outweigh wage gains from permanent migration \citep{Munshi2016}.
Within districts, \citet{Colmer2021} finds that temperature shocks \textit{reduce} 
demand for agricultural labour, increasing nonagricultural employment in the short run.

On the other hand, \citet{LST2023} document that higher temperatures \textit{raise} 
district-level agricultural labour shares, through reduced nonagricultural demand
in the long run.
Across countries, \citet{Nath2025} shows that despite heat-induced productivity losses,
a ``food problem'' intensifies agricultural specialisation in developing economies,
exacerbating the ``agricultural productivity gap''.\footnote{Agriculture continues 
to employ roughly 42\% of India's contemporary workforce (World Bank, 2025).}
The debate remains unsettled on whether sectoral reallocation of labour 
can mitigate local economic losses under a warming climate.

This paper offers one resolution to the debate on warming temperatures and structural transformation:
whether agricultural employment shares rise or fall follows the relative magnitudes of rising 
nighttime minima $(T_{min})$ or daytime maxima $(T_{max})$---compressing or expanding the diurnal temperature range (DTR).
The headline result is a divergence in agricultural wage-labour shares based on the
\textit{time of day} where warming is most prominent: warmer nights raise, warmer days lower this share.

This night-day divergence in agricultural labour is rooted in crop and human physiology: crops stand on land
around the clock, while farmers are exposed most under daylight. Consequently, warmer nights raise nocturnal
respiration, spending the carbon crops fix by day \citep{Peng2004, Welch2010}; warmer days press
field labour against a physiological ceiling, lowering the productivity of labour itself
\citep{Dunne2013, Somanathan2021}.
Agronomy has documented that nighttime and daytime warming carry distinct effects 
on crop yields and pull in different directions on overall production \citep{Peng2004, Lobell2007, Welch2010, Chen2016}.
The diurnal decomposition, however, has not crossed into the economic literature on structural transformation.
From a decadal panel and long-differenced
approach within Indian Census districts and towns (1981--2011), I find that higher nighttime ($T_{min}$)
temperatures \textit{hold} agricultural labour on land while higher daytime ($T_{max}$)
temperatures \textit{release} it---even though both measures reduce agricultural production and
land-use.

Prior economic literature has primarily measured warming through `degree days', a count of extreme days,
or binned daily temperatures \citep{Schlenker2006, Blakeslee2015, Dell2014, Somanathan2021}. These 
daily measures average temperatures over the diurnal cycle, and therefore collapse the night-day distinction 
that agronomy holds apart.
The closest neighbours to this paper---studying Indian labour reallocation---likewise treat 
warming as a scalar: \citet{Colmer2021} and \citet{LST2023} identify structural-transformation
effects averaging diurnal temperatures.
By separating the nighttime and daytime margins, this paper uncovers an additional dimension
to the relationship between temperature and agricultural productivity shocks
through factor-bias \citep{Bustos2016, BarrettOrtizBobeaPham2021}
under asymmetric diurnal warming.
An equivalent representation is the level and the diurnal range of surface-air temperatures.
Together, they separate two distinct mechanisms of structural transformation in the local economy.

\begin{figure}[!htbp]
    \centering
    \caption{The diurnal warming geography of India}
    \label{fig:warming_geography}
    \includegraphics[width=\linewidth]{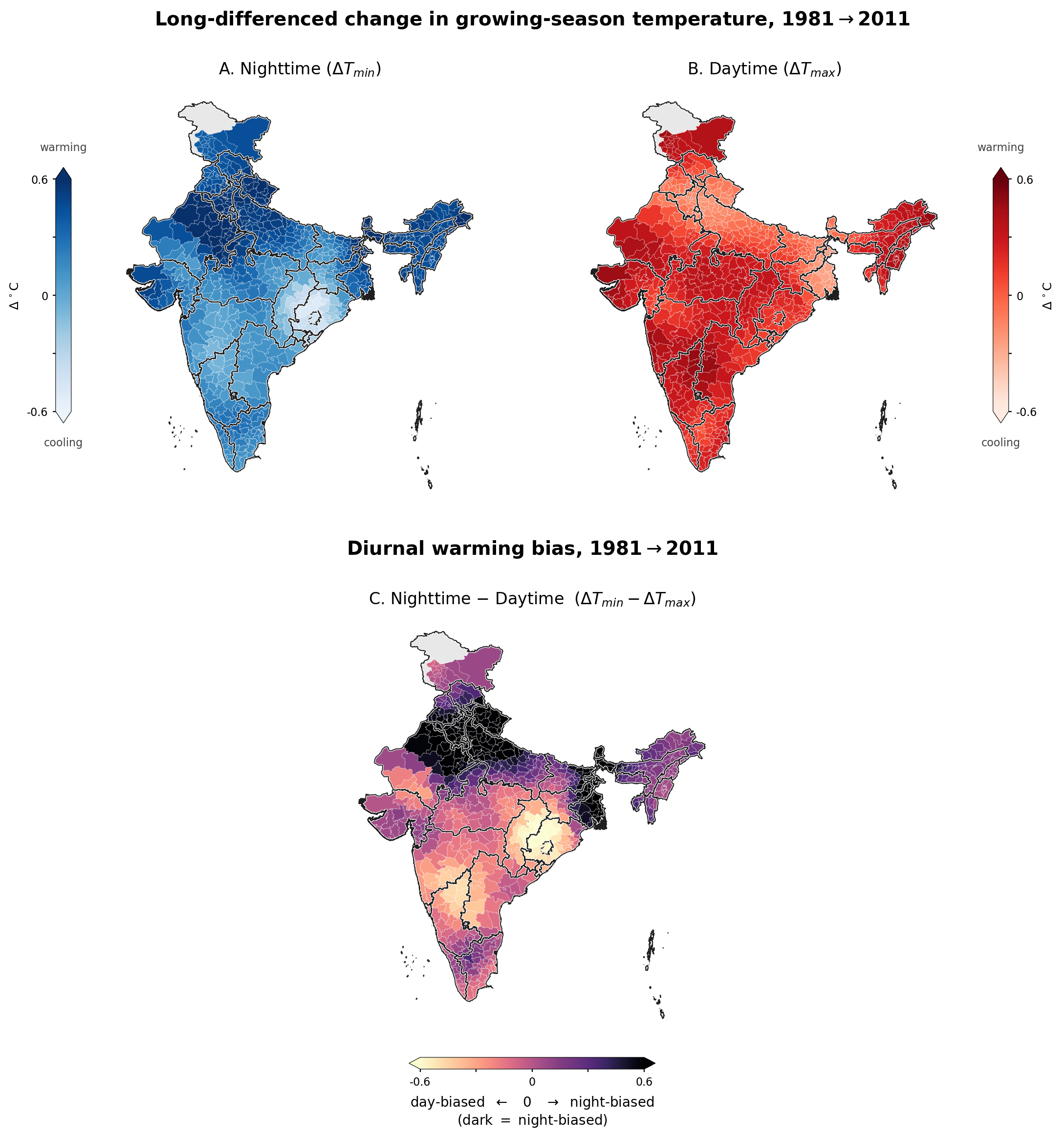}
    \parbox{\linewidth}{
        \vspace{0.5em}
        \footnotesize
        \textit{Notes:}
        Spatial distribution of warming nighttime and daytime temperatures
        in India.
        This map plots the change in decadal temperatures over the growing-season (June--Feb),
        by long-differencing the average between two endpoints: (2001,2011) $-$ (1981,1991).
        IMD gridded climate is assigned to each district centroid by inverse-distance weighting.
        Panels A and B map the change in the nighttime minimum ($\Delta T_{min}$) and the 
        change in daytime maximum ($\Delta T_{max}$)---darker where warming is stronger.
        Panel C maps the diurnal warming bias
        ($\Delta T_{min}-\Delta T_{max}$): darker where nights warmed faster (compressing
        DTR) and lighter where days warmed faster (widening DTR).
        Across the 639 districts with map coverage, nights outpaced days across the
        Indo-Gangetic north, while days outpaced nights
        across the peninsular Deccan (the estimation panel covers roughly 300 districts,
        Section~\ref{sec:data}).
    }
\end{figure}

Figure~\ref{fig:warming_geography} shows the asymmetric warming geography of India, obtained by
long-differencing the 10-year average of nighttime and daytime temperatures in districts 
between two endpoints (1981,91 -- 2001,11).
Although night and day temperature `levels' rose across the board, the bias on the DTR
sorts across space.
Asymmetric diurnal warming \citep{Karl1991, Karl1993, Easterling1997, Davy2016, IPCC_AR6} reached 
India after 1991 \citep{Mall2021}:
the DTR contracted in the arid northwest and Indo-Gangetic plains,
and expanded in the Deccan peninsula and along the coasts.
This geography provides continuous variation in nighttime versus daytime warming across
districts and towns.

This paper establishes three core empirical facts.

First, within districts, decadal panel estimates (Table~\ref{Table1}) show that higher
nighttime and daytime temperatures both reduce production and area under cultivation of 
grains, while only daytime temperatures raise their harvest prices locally.\footnote{The
core grain basket of India is composed of rice, wheat, maize, barley, sorghum and millets.}
Under both shocks, cultivated area falls with a similar 
magnitude ($\btmin=\btmax \approx -0.22$ log points, $F = 0.01$).
Notably, nighttime warming produces a nominally milder loss in grain \textit{production} 
$(-0.15)$ relative to daytime warming $(-0.22)$, with no detectable change in local harvest prices.
Daytime warming acts as a physical supply shock, spiking local harvest prices by $0.13$ log points,
rejecting equality with night warming $(F = 13.34)$.

Second, the direction of change in agricultural wage-labour shares reverses between night 
and day warming. A $1\degree$C decadal rise in nighttime
minimum temperatures $(\btmin)$ \textit{increases} agricultural wage-labour shares by $+4.4$
percentage points within districts, drawing from self-cultivators ($-2.8$) and seasonal
`marginal' workers ($-3.3$).\footnote{The
Census of India classifies `marginal workers' as those employed under 183 calendar days preceding 
the year of enumeration---interpreted as `seasonal workers'.}
An equivalent $1\degree$C decadal rise in the daytime maximum temperatures $(\btmax)$
\textit{decreases} agricultural wage-labour shares ($-2.6$) and raises seasonal worker
shares ($+3.2$).
The two coefficients reject joint equality $(\btmin=\btmax)$ on agricultural
labour shares ($F = 24.9$) and seasonal worker shares ($F = 14.7$). 

The opposing signs on agricultural labour shares are robust to humidity-adjusted 
wet-bulb temperatures ($+0.086 / -0.079$) and a long-differenced 
estimation ($\bdtmin=+0.075, \bdtmax=-0.062$). A unidimensional measure folds the 
opposing effects into a single positive coefficient---growing-season degree-days 
yield $\beta_{DDs}=+0.036$ on the same panel 
(online appendix)---because warming in this period is night-biased. 

Third, the diurnal reversal on agricultural labour shares scales with the weight of 
the agricultural sector in the local economy: over two-decade long differences, 
the night-day divergence is strongest among rural workers
$(F=30.9)$ and muted in the urban workforce $(F=1.9)$ (Figure~\ref{fig:ruralurban}).
Both rural and urban workforces share a secular decline $(\Delta \bar{y})$ over
1981--2011 ($\Delta \bar y^{rural}=-0.046$ rural, $\Delta \bar y^{urban} = -0.035$ urban).
Within towns (Figure~\ref{fig:towns}), where agriculture is too small to register the divergence,
both margins instead \emph{erase} the agricultural-labour decline to a net near-zero (no
night-day split, $F=1.61$) and jointly push down the non-agricultural share
($\bdtmax=-0.08$, $\bdtmin=-0.04$, $F=2.78$).

Over long differences, the district-level night-day rural divergence
presents a divide between nighttime warming consolidating rural agricultural labour into
year-round (main-status) wage work, and daytime warming returning rural labour to the casual
seasonal margin---a seasonal return that is sharp in the decadal panel ($+3.2$ points,
Table~\ref{Table2}) and directionally present over the long difference ($+0.056$, $t=1.58$).
Together, the two margins converge at the town-level to reduce non-agricultural
employment shares, consistent with a loss in labour productivity
and dampened demand for non-agricultural goods and services \citep{LST2023}.

Read together, the results present a two-pronged arrest of 
structural transformation: warmer nights consolidate agricultural 
labour, warmer days casualize it, and both cut agricultural production.

These empirical facts share a common root: nighttime ($\tmin$) and daytime warming ($\tmax$) reach the
farm through different factors---a warmer night shocks the land, while a warmer day is a
neutral productivity loss eroding labour and the underlying land productivity alike.
Section~\ref{sec:theory} synthesizes these findings under a conceptual framework, discussing
the assumptions and mechanisms. It maps the daytime warming supply shock to locally rising
harvest prices. The framework presents two testable implications---the price asymmetry, that
only daytime warming moves the harvest price, and the rural--urban amplitude ordering. An
online appendix formalises the full general-equilibrium model and presents robustness
decompositions across space, time, and measurement, verifying that the three central empirical
facts are robust to specification.

A final piece of suggestive evidence is a near-decadal panel of urban households from three
rounds of the National Sample Survey (1987, 1999, 2009) \citep{Ruggles2024}. These household
surveys corroborate the mechanism: warmer nights raise urban agricultural labour shares while
warmer days raise employment in trade, transport and logistics, mirroring the decadal
`seasonal' pattern in urban Censuses. With only three rounds of data, I read this as descriptive 
evidence. 

The distinction between the two temperature shocks is important for welfare.
The nighttime shock halts structural transformation by pulling more hands onto the farm;
the daytime shock diverts agricultural wage-employment into a low-productivity, 
informal churn within urban services.

This paper makes three contributions.
First, the diurnal decomposition sheds light on the conflicting short- and long-run estimates
of heat and labour reallocation in \citet{Colmer2021} and \citet{LST2023}: it shows what a
single degree-day average conceals, that warming nights sort labour within agriculture while
warming days reduce formal wage employment, so the direction of change depends on which end of
the day warms.
Second, a conceptual framework and general equilibrium model (online appendix) 
show the night/day divergence to be a property of factor bias under gross substitutability 
of labour and land, carrying an over-identifying 
restriction---only daytime warming moves the local harvest
price ($F=13.34$), which is consistent with partial market integration. 
Third, I construct a novel harmonisation of a four-decade panel (1981--2011) linking 
agricultural statistics (ICRISAT) with employment shares (Census, NSS) at the 
district and town level, covering a critical economic transition of the largest population
in the world.

The rest of the paper proceeds as follows.
Section~\ref{sec:theory} develops the framework.
Section~\ref{sec:data} describes the data and the empirical approach.
Section~\ref{sec:results} reports results. 
Section~\ref{sec:conclusion} concludes.

\section{Conceptual Framework}
\label{sec:theory}

This section describes the mechanism and economic channels mapping the night-day 
temperature margins to the labour reallocation estimated in the rest of the paper.
The full general-equilibrium model, with proofs, is in the online appendix.

\paragraph{What do nighttime and daytime temperatures measure?}
The nighttime minimum and the daytime maximum are the floor and the ceiling of the daily
temperature cycle. Holding one fixed, a higher minimum compresses the diurnal range and a
higher maximum widens it. 
The two margins act on different inputs in the production function, which are compressed 
into one heat shock when measured as daily averages.
Crops occupy the field around the clock, so a warmer night works on the
standing crop and the land in cultivation, raising respiration and the overnight heat load
the crop carries \citep{Peng2004, Welch2010}. Field labour occupies only the daylight, so a
warmer day works on the worker, pressing effort against the physiological ceiling of
sustained outdoor exertion and lowering the productivity of labour itself
\citep{Dunne2013, Somanathan2021}. 
A joint specification of the minimum and maximum temperatures separates the
daily \emph{level} (their mean) from the diurnal \emph{range} (their difference): the level
moves the farm's scale---area, output, and self-cultivation---while the range drives the
reallocation between wage labour and the seasonal margin and the harvest price.

\paragraph{Nighttime temperature factor-bias in agricultural production.}
Agricultural production combines labour and land. Nighttime warming acts a
\emph{land-biased} shock that raises or lowers the productive services of cultivated land,
while daytime warming is a \emph{Hicks-neutral} loss of total factor productivity.
Under factor-bias, the night margin distinguishes endogenous \emph{labour adaptation}.
In principle, the nocturnal-respiration and overnight-heat-load channels can either
degrade or enhance effective land. The data disciplines the sign of this change
empirically---higher wet-bulb night temperatures \emph{raise} cereal yields, while 
dry-bulb night temperatures have no detectable efffect on yields.

One central assumption is that labour and land are gross substitutes with 
an elasticity of substitution $\sigma>1$, which is not directly estimated in this design. 
This substitutability assumption
is grounded in prior literature under similar agricultural contexts 
\citep{HayamiRuttan1985,Binswanger1974,BinswangerRosenzweig1986}.
Cross-country CES estimates cluster around $\sigma=1$, varying with method and setting \citep{Mundlak2012}. 

I maintain $\sigma>1$ as the empirically relevant case (online appendix).
The empirical patterns discipline this assumption: agricultural labour rise under warmer nights,
while area and output contract and the price stays flat. 
When a warmer night
\emph{augments} the quality of an acre, a farm that substitutes readily reworks its
land-to-labour mix and works the land harder, hiring \emph{more} field labour rather than
less; a degrading night shock would instead lower labour demand. The sign of the shock to the
factor sets the direction of the labour response, independent of yields---consistent with findings  
from \citet{Bustos2016} in Brazil, where land-augmenting maize pulled labour into agriculture and
labour-saving soy released it. 

In this paper, the diurnal heat cycle supplies the two biases.
A neutral productivity loss lands on no factor in particular: it leaves the
land-to-labour ratio intact and shrinks the scale of the operation, pulling down the marginal
product of labour and the field wage and, where the food market is tight, raising the harvest
price. The two margins therefore move farm-labour demand in \emph{opposite} directions---warmer
nights raise it, warmer days lower it. Under Cobb--Douglas
($\sigma=1$) a land-augmenting shock folds into total factor productivity, the two channels
collapse, and a daily average loses nothing.

\paragraph{The seasonal employment `sink'.}
Workers divide across four states---agricultural wage labour, self-cultivation,
non-agricultural work, and casual seasonal work---whose shares sum to one, so the margins
move as one offsetting system, and the casual seasonal pool is the flexible residual that
clears the reallocation. When nighttime warming raises the demand for farm labour, the farm
hires: the wage-labour share rises and draws from the cultivators who give up their marginal
plots and the casual workers who attach to the harvest---the mark of a commercialising farm
that retains and reorganises its labour. When daytime warming collapses
the marginal product, the wage-labour market gives way and returns workers to the seasonal
margin. Self-cultivation falls under both margins, as marginal plots go unworked.

\paragraph{The price test for market integration.}
I test the decomposition's openness rather than assume it. Only a productivity-driven (daytime)
output loss raises the local harvest price; the nighttime contraction leaves it flat. That
asymmetry marks a district whose food price is set largely by trade, not
by local scarcity (partial market integration). The online appendix embeds this in a
general-equilibrium closure spanning open and closed food markets and shows that the
over-identifying price restriction places India's districts on the
integrated side; a fully closed subsistence economy would instead invert the labour
orientation. Partial integration lets a daytime output loss raise the local harvest
price, which is why only the daytime margin moves it.

\paragraph{Intepreting the rural-urban channels.}
Rural and urban populations share the nighttime sign and differ in amplitude; the urban
daytime response is a statistical zero.
The factor-bias divergence is a mechanism that operates inside the agricultural sector;
its footprint in local employment shares therefore scales directly with that sector's
weight in the local economy.
Where agriculture is the dominant sector---rural district populations---the reallocation
registers sharply, amplified further by the depth of the casual-labour margin that absorbs it.
Where agriculture is a minor sector---towns---the within-agriculture divergence attenuates,
and the dominant margin becomes the sector's own heat response: a shared loss of capacity for
outdoor informal work that contracts non-agricultural employment under both margins. This is
consistent with the framework's supply channel and with dampened local demand for non-farm
goods and services under heat \citep{LST2023, Nath2025}.
The framework belongs to the structural-transformation tradition
\citep{Matsuyama1992Dec, Kongsamut2001Oct, Gollin2014, Bustos2016, BarrettOrtizBobeaPham2021};
the online appendix develops the full general-equilibrium model---primitives, the
gross-substitutes assumption, the price closure, and the lemma and propositions behind the
predicted signs, with all proofs. These signs are meant as guidelines to reading
the direction of change in our variables of interest.

\section{Data and Empirical Approach}
\label{sec:data}

\subsection{Data Sources}

I harmonise four datasets to link climate, agricultural production and structural transformation: (1) daily gridded weather data from the IMD, (2) district agricultural data (ICRISAT), (3) district and town level tables from the Indian Census, and (4) urban household-level employment data from the NSS.
These datasets span four decades (Census, IMD, ICRISAT), with a shorter window from the NSS (1987, 1999, 2009).

\paragraph{Indian Meteorological Department}
Gridded data were obtained from the Indian Meteorological Department (IMD) on air temperature ($1\degree$ resolution) and precipitation ($0.25\degree$ resolution) over 1951--2023.
Consistent with the approach in \citet{LST2023}, daily observations are averaged over growing seasons of the past 10 years.
The growing season marks the months of June--February, which captures cropping cycles of rice in the monsoon (kharif) and wheat in the winter (rabi) seasons.
The baseline exposure measures are the growing-season means of the daily minimum and
maximum surface-air temperatures, $T_{min}$ and $T_{max}$, which separate nighttime
from daytime warming. For comparison I also construct degree days (DDs), a scalar that
rescales and sums agriculturally relevant temperatures between $8$ and $32\,\degree$C
\citep{Blakeslee2015, Schlenker2006}; degree days average over the diurnal cycle and so
collapse the two margins into one.
    \begin{equation}
        DDs = \sum_d D(\text{T}_{d}) \text{ ; where }
        \text{T}_{d} = \frac{\text{T}_{min,d} + \text{T}_{max,d}}{2}
    \end{equation}
    \begin{equation}
        \label{eq:dds_calc}
        \begin{aligned}
        D(\text{T}) &=
        \begin{cases}
            0, & \text{if } \text{T} \leq 8^\circ\text{C} \\
            \text{T}-8, & \text{if } 8^\circ\text{C} < \text{T} \leq 32^\circ\text{C} \\
            24, & \text{if } \text{T} > 32^\circ\text{C}
        \end{cases}
        \end{aligned}
    \end{equation}

As a humidity-adjusted robustness measure I also construct growing-season wet-bulb minimum
and maximum temperatures following \citet{Stull2011Nov}; these are derived from $T_{min}$ and
$T_{max}$ alone---no external humidity series is used (the construction is detailed in the
supplemental data appendix)---and enter the wet-bulb specifications reported there.

\paragraph{Census of India}
The core dataset is sourced from the District Census Handbooks (DCHBs) of India.
Specifically, I digitise the Primary Census Abstract (PCA) tables, covering the universe
of Indian towns from 1981 to 2011.\footnote{The town digitisation builds on data assembled
for \citet{BlakesleeDar2023}.} A Census town meets three criteria: (1) a minimum
population of 4,000, (2) at least 75\% of the male working population in
non-agricultural activity, and (3) a population density above 400 persons per sq.\ km.
I construct a panel of 2{,}969 towns that appear consistently across all four decades,
so that the analysis reflects intensive-margin adjustments rather than the
reclassification of urban areas. For comparability over time, occupations are
aggregated into four mutually exclusive categories: cultivators (landowners),
agricultural labourers (wage workers), non-agricultural workers, and marginal workers.
The ``marginal'' category cannot be assigned cleanly to agriculture or
non-agriculture, but it lets me track the contractual \textit{quality} of work.
The 2001 Census enumerated marginal work more completely than 1991; the headline
results are identified off the consistently defined main-worker categories, and a
per-capita denomination immune to the change reproduces them (online appendix).
The harmonised district panel covers roughly 300 districts across twenty major states
on 2011-consistent boundaries; eight north-eastern states lack a usable 1981
identifier and are excluded (online appendix).\footnote{Of the 300 districts on
2011-consistent boundaries, 299 admit a computable long difference in the agricultural
wage-labour share. The regressions absorb state fixed effects, dropping six single-district
states and union territories as singletons (Chandigarh, Delhi, Dadra \& Nagar Haveli,
Lakshadweep, Puducherry, Andaman \& Nicobar) for $N=293$. The residence-specific panels lose
two more each---two fully urban districts (Mumbai Suburban, Chennai) have no rural population
and two fully rural ones (Lahul \& Spiti, Kinnaur) no urban---so that share is undefined
($N=291$); the log-worker columns retain $293$.}

\paragraph{ICRISAT}
The ICRISAT District Level Database (DLD) tracks agricultural inputs and production in India from 1966 to the present.
I harmonise districts to 2011-Census consistent boundaries, constructing an annual panel of cultivated area, production, yields and harvest prices for key grains: rice, wheat, maize, sorghum, barley and millets.
I construct the mean of these measures over 10 years preceding each Census round, e.g. over the years 1971--1980 for the 1981 Census.
Thus, the results from the temperature regressions compare contemporaneous horizons of 10-year windows with climate data.
The DLD covers staple grains only; horticulture and cash crops---which have grown rapidly and differ in labour intensity, irrigation, and market structure---lie outside it, so the agricultural-economy results speak to staple-grain production and the broader composition of Indian agriculture is left to future work. The Census labour reallocation, however, survives this composition shift: it is strongest in districts that stayed in staple grains and muted where agriculture diversified into high-value crops (online appendix).

\paragraph{National Sample Surveys}
I use three rounds of the National Sample Survey (NSS)
Employment and Unemployment Schedules (1987, 1999, and 2009), harmonised by IPUMS
\citep{Ruggles2024}.
While the Census measures the volume of labour reallocation, it lacks a granular 
view of the destination sectors within non-agriculture under its 1981 definitions. 
Restricting the NSS sample to urban households serves as an independent benchmark to the
Census urban and town-level data. The NSS urban sampling frame is separate
from Census towns (it retains more peri-urban agriculture), and provides complementary
evidence of the divergence in agricultural wage-labour shares. The data additionally shed
light on the destination of heat-displaced labour---casual, informal work rather than stable
formal employment in manufacturing or services.

\subsection{Empirical Approach}

The identification takes two approaches: a fixed effects panel estimation (\ref{eq:eq_panel}), and a long differences approach (\ref{eq:eq_ldiff}), consistent with \citet{Dell2014}, \citet{BurkeEmerick2016} and \citet{LST2023}.
Let $\textit{Y}_{irt}$ be the outcome of interest (e.g., share of workers in agriculture) for district or town $i$, within state $r$, observed at time $t$.
$\mathbf{T}_{irt}$ denotes a vector of temperature measures, either degree days
$\mathbf{T} = \{DDs\}$ or the diurnal surface-air pair
$\mathbf{T} = \{T_{min}, T_{max}\}$. All specifications control for total precipitation
$\mathbf{P}_{it}$ and fit a state-specific linear time trend.

\begin{equation}
    Y_{irt} = \beta \mathbf{T}_{irt} + \phi \mathbf{P}_{irt} + \mu_{i} + \lambda_{t} + \gamma_{r} \cdot t + \varepsilon_{irt}
    \label{eq:eq_panel}
\end{equation}

\begin{equation}
    \Delta Y_{ir} = \beta {\Delta \mathbf{T}}_{ir} + \phi \Delta \mathbf{P}_{ir} + \gamma_{r} + \varepsilon_{ir}
    \label{eq:eq_ldiff}
\end{equation}

For temporal parity, I construct the gridded temperature and rainfall measures as a
moving average over the growing seasons of the preceding ten years. Decadal climate is
the inverse-distance-squared--weighted average of IMD grid points surrounding the
district (within 200 km) or town (within 100 km) centroid---radii chosen to match the
distance at which the residual spatial correlation of the outcomes decays (roughly
150--200 km for districts and 100 km for towns). Because the weights fall off
as the squared distance, grid points beyond roughly 100 km contribute negligibly, so the
assignment is insensitive to the exact cutoff.

Equation \ref{eq:eq_panel} is a district- or town-decade panel with two-way fixed effects and a
state-specific linear trend; because the climate regressor is itself a ten-year moving
average, I read its coefficients as the \textit{decadal} response of structural transformation
to heat. Equation \ref{eq:eq_ldiff} averages the outcome $Y$ and covariates $X$ over two
endpoints---1981--91 and 2001--11---and regresses $\Delta Y = Y_2 - Y_1$ on
$\Delta X = X_2 - X_1$. This long-differenced estimator recovers the effect of a sustained
increase in $\mathbf{T}$ on the outcome trend $\Delta Y$, measured against its secular trend
$\Delta \bar Y$ over the sample. Both estimators use the same decadally smoothed climate; the
long difference adds one further decade of horizon and absorbs unobserved state-level
trajectories $(\gamma_r \cdot t)$. I treat it as a robustness design---confirming that the
decadal-panel result survives a different fixed-effects and averaging structure that strips
decade-common shocks---rather than as a distinct long-run or adaptation-inclusive
estimand.\footnote{\citet{Lemoine2021} shows that long-difference and panel estimators of
climate impacts are identified by overlapping variation in the realised temperature series
(here decadally averaged), so a long difference need not isolate differential rates of
climate change or a separate long-run adaptation response, and agreement between the two
designs reflects robustness to the estimating structure rather than evidence on adaptation
dynamics. Identification here therefore rests on the within-unit, simultaneous
$T_{min}$-versus-$T_{max}$ contrast in the decadal panel; the ten-year smoothing and the long
difference serve as robustness rather than as a short-run/long-run decomposition.}

I enter $T_{min}$ and $T_{max}$ simultaneously and read each coefficient as the partial
effect of one margin holding the other---and rainfall---constant. Inference faces two distinct
dependencies. The first is the decadal \emph{persistence} of the climate regressor (a
ten-year moving average), which inflates rejection rates unless the estimator accounts for
within-unit serial correlation. This choice is discplined by Monte Carlo simulations (online 
appendix), where I regress the key outcomes on a placebo climate series---the real series 
randomly permuted across districts--- preserving decadal persistence.
I select the preferred panel estimator for equation
\ref{eq:eq_panel} as the specification that rejects the null in roughly 5\% of simulations.
I find that unadjusted and heteroskedasticity-robust standard errors over-reject sharply, whereas
clustering by district with state-specific linear trends restores the nominal 5\% size. 

The second dependency is spatial---a persistent, strongly associated feature of Indian
diurnal-temperature data \citep{BhattacharjeeBose2024}; because the permutation reassigns
whole series across units it breaks the regressor's spatial structure, so the Monte Carlo
does not speak to spatial correlation, which I address separately with Conley spatial-HAC
standard errors \citep{Conley1999} reported alongside the clustered errors on the panel
exhibits. The online appendix sweeps the Conley radius to 200~km: the night/day equality
rejection is stable at every radius ($F=12.4$ at 200~km), with the daytime labour
coefficient thinning to the ten-percent level at the widest radius.

The simultaneous estimator addresses two identification challenges. First, because
$T_{min}$ and $T_{max}$ are correlated ($\rho \approx 0.78$ across districts), entering
either alone would induce omitted-variable bias; entering them together isolates their
partial effects and identifies a narrowing versus widening of the local diurnal range. The
district and decade fixed effects and state trends absorb most of this co-movement, so the
\emph{estimating} variation is close to orthogonal (within-transformed
$\mathrm{VIF}\approx 1$; the online appendix) and the partial signs are
well-identified. By Frisch--Waugh--Lovell, the coefficients are determined exactly from this 
residual variation, so the within correlation governs whether the two
margins can be separated---and within, they are all but orthogonal. Absorbing state-specific 
trends accounts for time-varying state policies and compares
districts and towns within their local economic geography under similar baseline climate
trajectories. Correlograms of the residuals confirm that the spatial correlation of
these heat shocks in the key outcomes decays within roughly 100 km for towns and 150--200 km
for districts.

\section{Results}
\label{sec:results}

\begin{table}[!htbp]
    \footnotesize
    \centering
    \caption{Panel estimates of diurnal warming on the district agricultural economy}
    \label{Table1}
    \renewcommand{\arraystretch}{1.2}
\begin{tabular}{l*{5}{c}}
\toprule
& \multicolumn{5}{c}{Log Levels (All Grains)} \\
\cmidrule(lr){2-6}
& \shortstack{Cultivated \\ Area} & \shortstack{Production} & \shortstack{Harvest \\ Price Index} & \shortstack{Field \\ Wage} & \shortstack{Yield \\ Index} \\
& (1) & (2) & (3) & (4) & (5) \\
\addlinespace
 \\
$ T_{min}$ & \textbf{-0.218} & \textbf{-0.145} & -0.042 & -0.038 & 0.051 \\
\addlinespace[4pt]
   & (-4.76) & (-2.26) & (-1.36) & (-0.62) & (1.27) \\
   & [-4.76] & [-2.25] & [-1.36] & [-0.63] & [1.28] \\
\addlinespace[8pt]
$ T_{max}$ & \textbf{-0.225} & \textbf{-0.221} & \textbf{0.130} & -0.119 & -0.046 \\
\addlinespace[4pt]
   & (-4.01) & (-3.47) & (3.81) & (-1.70) & (-0.97) \\
   & [-4.02] & [-3.46] & [3.77] & [-1.76] & [-0.97] \\
\addlinespace
  \\
Observations & 1,243 & 1,243 & 1,191 & 1,115 & 1,211 \\
Dep Var. Mean & 5.45 & 5.77 & 5.66 & 2.86 & 7.14 \\
F-stat ($ T_{min}=T_{max}$) & 0.01 & 0.80 & 13.34 & 0.74 & 2.63 \\
\quad \textit{p}-value & 0.905 & 0.373 & $<$0.001 & 0.391 & 0.106 \\
\midrule
\addlinespace
Rainfall Control & Y & Y & Y & Y & Y \\
State Linear Trend & Y & Y & Y & Y & Y \\
\bottomrule
\end{tabular}

    \parbox{\linewidth}{
        \vspace{0.5em}
        \textit{Notes:}
        Log cultivated area, log production, a log harvest-price index, the log field wage,
        and an area-weighted yield index for the major Indian cereals (rice, wheat, barley,
        maize, millets, sorghum), from the ICRISAT district panel. $T_{min}$ and $T_{max}$ are
        the growing-season (June--February) surface-air minimum and maximum, averaged over the
        preceding decade; all specifications control for total precipitation over the same
        window. The data form a decadal panel (1981--2011) on 2011 Census-consistent district
        boundaries, with district and decade two-way fixed effects and a state $\times$ decade
        linear trend. Parentheses report $t$-statistics from district-clustered standard
        errors; brackets report $t$-statistics from Conley spatial--temporal HAC standard
        errors (50~km, four decadal lags); bold coefficients indicate $|t|>2$. The
        $F$-statistic and its $p$-value test $T_{min}=T_{max}$.
    }
\end{table}

\begin{table}[!htbp]
    \footnotesize
    \centering
    \caption{Panel estimates of diurnal warming on district labour reallocation}
    \label{Table2}
    \renewcommand{\arraystretch}{1.2}
\begin{tabular}{l*{5}{c}}
\toprule
& \multicolumn{4}{c}{Share of Workers} & \\
\cmidrule(lr){2-5}
& \shortstack{Ag. \\ Labour} & \shortstack{Cultivator} & \shortstack{Seasonal} & \shortstack{Non- \\ Agriculture} & \shortstack{Log \\ Workers} \\
& (1) & (2) & (3) & (4) & (5) \\
\addlinespace
 \\
$ T_{min}$ & \textbf{0.044} & \textbf{-0.028} & \textbf{-0.033} & 0.017 & 0.175 \\
\addlinespace[4pt]
   & (4.59) & (-2.21) & (-2.83) & (1.46) & (1.99) \\
   & [4.65] & [-2.23] & [-2.83] & [1.48] & [2.02] \\
\addlinespace[8pt]
$ T_{max}$ & \textbf{-0.026} & -0.010 & \textbf{0.032} & 0.004 & -0.017 \\
\addlinespace[4pt]
   & (-2.76) & (-0.83) & (2.45) & (0.32) & (-0.20) \\
   & [-2.77] & [-0.84] & [2.43] & [0.32] & [-0.21] \\
\addlinespace
  \\
Observations & 1,187 & 1,187 & 1,187 & 1,187 & 1,187 \\
Dep Var. Mean & 0.18 & 0.32 & 0.17 & 0.32 & 13.45 \\
F-stat ($ T_{min}=T_{max}$) & 24.91 & 0.85 & 14.66 & 0.54 & 2.36 \\
\quad \textit{p}-value & $<$0.001 & 0.356 & $<$0.001 & 0.462 & 0.125 \\
\midrule
\addlinespace
Rainfall Control & Y & Y & Y & Y & Y \\
State Linear Trend & Y & Y & Y & Y & Y \\
\bottomrule
\end{tabular}

    \parbox{\linewidth}{
        \vspace{0.5em}
        \textit{Notes:}
        District shares of total workers by category---agricultural wage labour, cultivators,
        the casual seasonal margin, and non-agriculture---and, in the last column, the log
        number of district workers, from the Census district panel. Outcomes, sample, fixed
        effects, and inference follow Table~\ref{Table1}: a decadal panel (1981--2011) with
        district and decade fixed effects and a state $\times$ decade linear trend, a
        precipitation control, district-clustered $t$-statistics in parentheses over Conley
        spatial--temporal HAC $t$-statistics in brackets, and bold coefficients for $|t|>2$.
        The $F$-statistic and its $p$-value test $T_{min}=T_{max}$.
    }
\end{table}

Tables~\ref{Table1} and~\ref{Table2} decompose the effects of warming nights and days on the
district agricultural economy: Table~\ref{Table1} reads down the supply chain for the major
cereals---rice, wheat, maize, sorghum, barley, and millets---and Table~\ref{Table2} reports the
within-district reallocation of labour. The nighttime minimum and the daytime maximum carry
opposite signs across the price and labour outcomes, consistent with the theoretical framework.

Both margins contract the farm. Cultivated land falls by a statistically identical amount
across the two margins (a $1\degree$C rise cuts log area by 0.22 either way, $F=0.01$), while
production falls by comparable amounts ($-0.15$ at night, $-0.22$ by day, $F=0.8$). I read the
milder nighttime loss as labour intensification; the significant nighttime rise in wage labour
adds output back through the production identity, cushioning the extensive-area loss, while
daytime labour-shedding compounds it. The two margins diverge on the harvest price. The daytime
maximum raises the log harvest price by 0.13 while the nighttime minimum holds it flat, a split
sharp at $F=13.34$; the field wage edges down under both ($-0.12$ by day, $-0.04$ at night, too
close to separate). Daytime warming thus acts as a supply shock, the loss of land and grain
passing into a dearer harvest price; the framework reads the nighttime contraction as
concentrated on marginal land whose output is more nearly consumed on-farm, passing little
into the marketed surplus that sets that price (online appendix). Cereal yields move little
under either margin in the dry-bulb baseline; the augmentation the labour response implies
surfaces only under humidity-adjusted temperatures, where night yields rise outright
(online appendix). The aggregate also blends
opposing grain-level responses, with heat-tolerant coarse cereals gaining under warmer nights
while wheat loses, so the cereal index is a crop-composition average (online appendix).

Table~\ref{Table2} reports the labour reallocation. A warmer night raises the agricultural
wage-labour share by 4.4 points and draws that labour off self-cultivation ($-2.8$) and the
casual seasonal margin ($-3.3$); a warmer day reverses the flow, cutting the wage-labour share
by 2.6 points and returning 3.2 points to the seasonal margin. The two margins separate sharply
($F=24.9$ on wage labour, $F=14.7$ on the seasonal share), while non-agricultural employment
holds flat under both ($F=0.5$), so the district reshuffles labour within agriculture. The
extensive count rises at night, a warmer minimum lifting the log number of district workers by
0.18, at the margin of significance. The appendix framework grounds the split: nighttime warming acts on land as a land-biased
shock, and because labour and land are gross substitutes it raises the demand for field labour
directly, while daytime warming is a neutral productivity loss that shrinks the operation and
sheds labour to the seasonal pool.

\begin{figure}[!htbp]
    \centering
    \caption{Long-differenced estimates of diurnal warming on agricultural labour trends}
    \label{fig:ruralurban}
    \includegraphics[width=\linewidth]{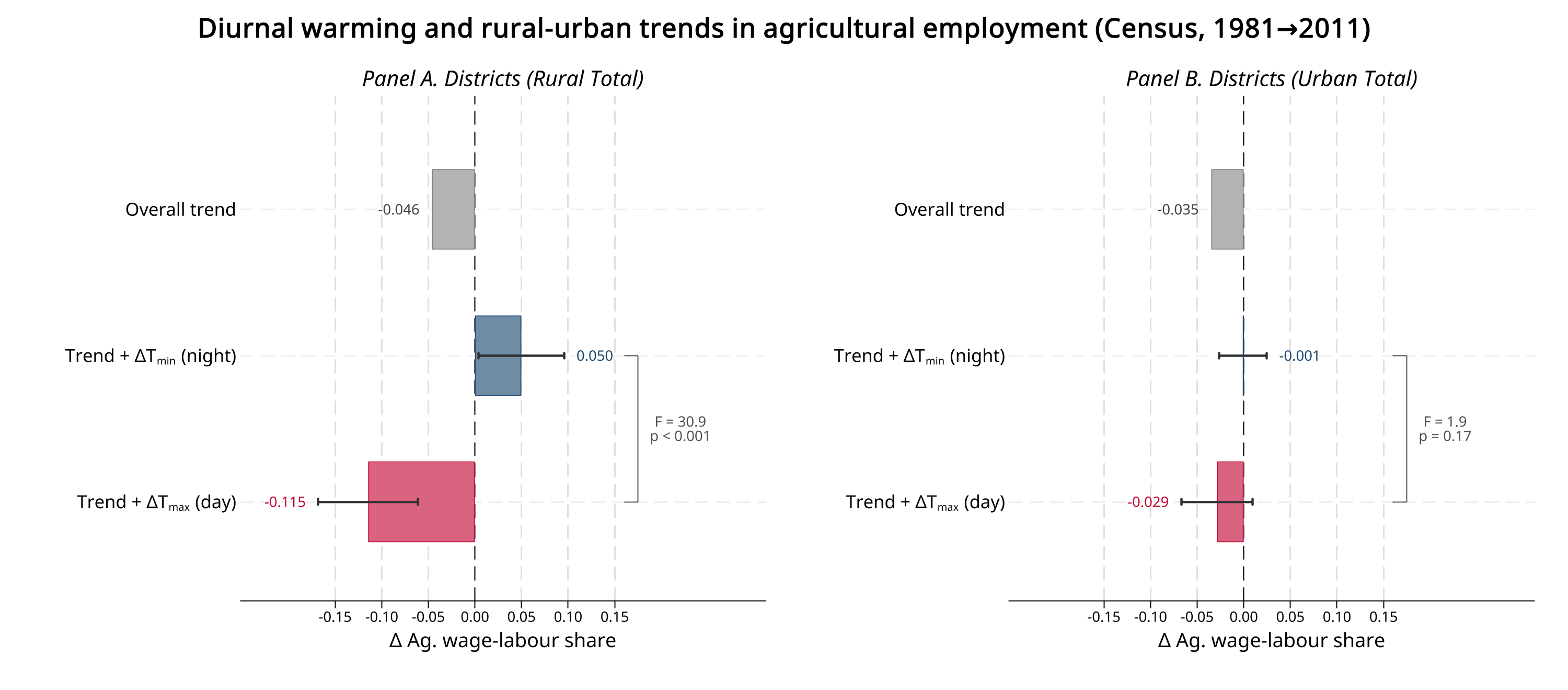}
    \parbox{\linewidth}{
        \vspace{0.5em}
        \footnotesize
        \textit{Notes:}
        Long-difference estimates (1981--2011) of a $1\degree$C rise in growing-season
        nighttime ($T_{min}$) and daytime ($T_{max}$) surface-air temperature on the district
        agricultural wage-labour share, estimated separately for the rural and urban populations
        of each district. Grey bars give the secular trend (the mean change in the share); each
        coloured bar adds the warming coefficient to that trend ($\bar{y}+\hat{\beta}$), with
        95\% confidence intervals (spikes); the brace reports the night-versus-day equality
        test ($\Delta T_{min}=\Delta T_{max}$). A coloured bar whose interval clears its grey
        trend bar is a statistically significant warming effect. All specifications control for the long-run
        change in precipitation and include state fixed effects; standard errors are clustered
        at the district level. The full total/rural/urban coefficients behind this figure are in
        Appendix Table~\ref{tab:apx_distLD}.
    }
\end{figure}

Figure~\ref{fig:ruralurban} carries the labour decomposition to its two-decade horizon and
splits each district into its rural and urban populations. The long difference averages each
district's two Census endpoints---1981 and 1991 against 2001 and 2011---and differences them,
stripping decade-specific noise to isolate the change in climate against the change in labour.
The grey bars set the stage: the agricultural wage-labour share fell across rural and urban
India over these two decades, the secular structural transformation drawing labour off the
land.

Warming bends that trend, and the two margins bend it in opposite directions. A warmer night
raises the rural agricultural wage-labour share by 0.096 and turns its $-0.046$ secular decline
into a $+0.050$ rise: nighttime warming hoards labour on the farm against the structural trend.
A warmer day lowers the same share by 0.068, deepening the decline to $-0.114$: daytime heat
pushes a rural agricultural exit. The opposition is sharp and holds over the two-decade horizon.

The urban populations of the same districts barely move. A warmer night lifts the urban
agricultural share by 0.034---a third of the rural response---and a warmer day leaves it flat
($+0.006$, insignificant). The reversal is a rural phenomenon; the urban populations of these
districts absorb little of it. The framework reads this
as amplitude (online appendix): the diurnal response scales with the
agricultural share of local employment, deep in rural districts and thin in urban ones, so the
same factor-bias force registers sharply in rural areas and faintly in the town.
The online appendix reports the full rural and urban decomposition behind the
figure---all four worker shares and log total workers. District populations move in the same
direction as workers, with the urban population rising under nighttime warming at the
ten-percent level (online appendix).

\begin{figure}[!htbp]
    \centering
    \caption{Diurnal warming and town labour reallocation over the long difference}
    \label{fig:towns}
    \includegraphics[width=\linewidth]{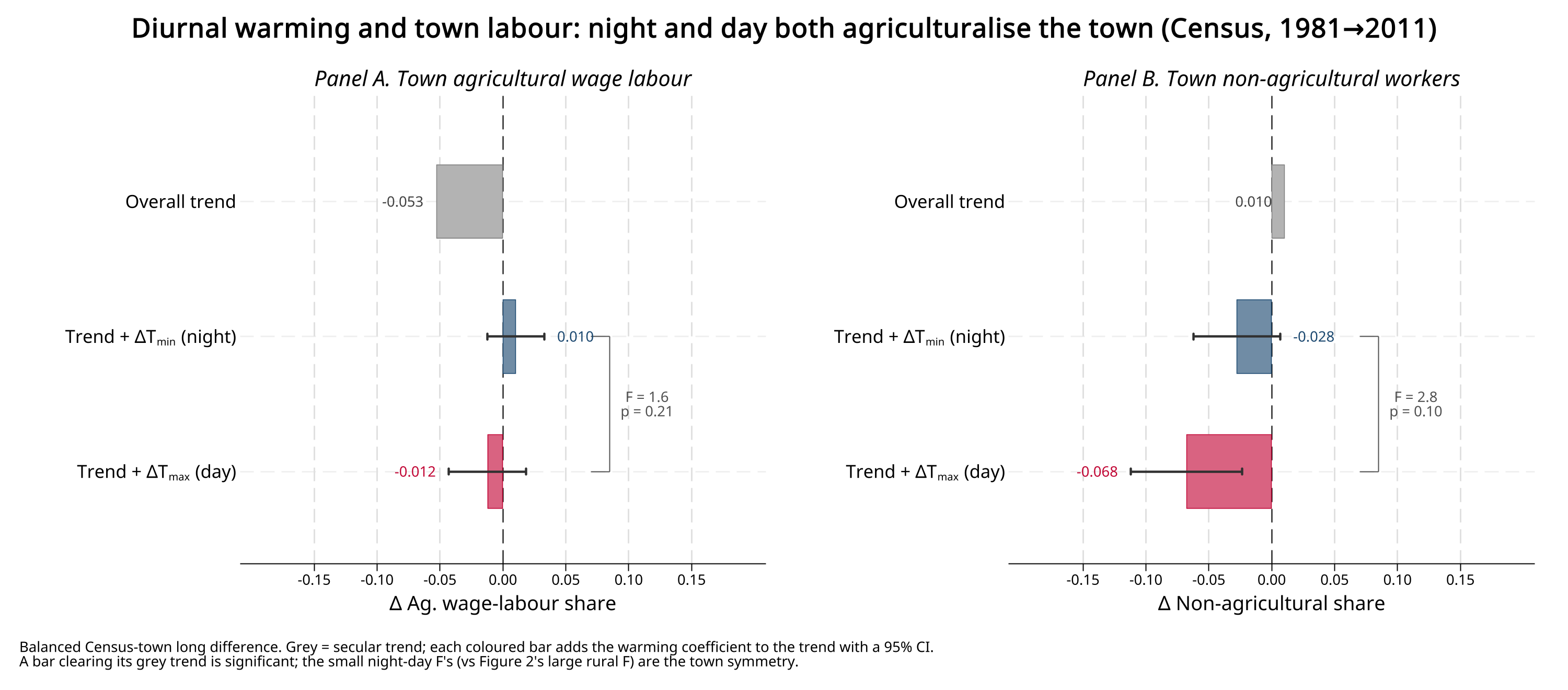}
    \parbox{\linewidth}{
        \vspace{0.5em}
        \footnotesize
        \textit{Notes:}
        Long-difference (1981--2011) estimates of a $1\degree$C rise in growing-season
        nighttime ($\Delta T_{min}$) and daytime ($\Delta T_{max}$) surface-air temperature on
        Census-town worker shares, for the balanced panel of towns. Panel~A is the agricultural
        wage-labour share; Panel~B the non-agricultural share. Grey bars give the secular trend
        (the mean change in the share); each coloured bar adds the warming coefficient to that
        trend ($\bar y+\hat\beta$), with 95\% confidence intervals; the brace reports the
        night-versus-day equality test ($\Delta T_{min}=\Delta T_{max}$). A coloured bar whose
        interval clears its grey trend bar is a statistically significant warming effect,
        read against Figure~\ref{fig:ruralurban}'s rural asymmetry. All specifications control
        for the long-run change in precipitation and include state fixed effects; standard
        errors are clustered at the district level. The underlying coefficients---all four
        worker shares and log town workers---are in Appendix Table~\ref{tab:apx_townLD}.
    }
\end{figure}

Figure~\ref{fig:towns} carries the decomposition to the Census town and closes the loop on the
central finding. Where agriculture is the dominant sector---rural district populations---the two
margins split sharply (Figure~\ref{fig:ruralurban}). Where it is a minor one---the urban
populations of those same districts, and the towns here---the wage-labour split collapses, and
warming does something else. It does two things at once. First, both margins \emph{erase} the town
agricultural-labour decline: the $-0.053$ secular fall is offset to a net near-zero ($+0.010$
under warmer nights, $-0.012$ under warmer days), with no day/night divergence on the wage-labour
share ($F=1.6$)---though the seasonal margin retains a residual night--day split ($F=4.1$,
$p=0.04$; Appendix Table~\ref{tab:apx_townLD}), a faint echo of the district pattern. This
is the same net erasure the urban district shows in Figure~\ref{fig:ruralurban}, where nighttime
warming lifts the $-0.035$ urban decline to $-0.001$: the positive nighttime coefficients do not
push the town toward agriculture so much as cancel its retreat from it. Second---and this is
what the town, unlike the urban district, makes visible---the non-agricultural share, which
drifted \emph{up} by about a point over these decades, is pushed down by both margins: warmer
days cut it sharply ($-0.078$) and warmer nights more mildly ($-0.038$), too alike to separate
($F=2.8$). The two margins move it together, so the daily \emph{level} and the diurnal
\emph{range} of warming both contract town non-agricultural employment. This town
de-urbanisation---the erased farm-labour decline alongside a squeezed non-farm sector---mirrors
\citet{BlakesleeDar2023}, where irrigation likewise raises rural agriculture and contracts the
town non-farm sector. The pattern is directionally the same in the full (unbalanced) town panel,
where the non-agricultural drag stays symmetric across the two margins though muted, alongside a
residual nighttime pull onto town agriculture; the balanced long difference, robustness across
the town sample, and the suggestive (imprecise) in-migration are in the online appendix.

A final, suggestive question is where the displaced labour lands. Matched to three rounds of the
National Sample Survey, the urban data reproduce the district mechanism on its matched
margin---a warmer night raises the urban agricultural wage-labour share ($+0.060$, $p<0.01$)---and
hint that daytime warming diverts labour into informal trade and transport rather than formal
employment (online appendix). With only three survey rounds I read this as
suggestive and leave it to future work.

The online appendix confirms the decomposition under alternative measures: a single
degree-days measure collapses the opposition into one muted coefficient; and wet-bulb
temperatures sharpen the labour split while humid nights raise cereal yields outright
($+0.15$), the land-augmenting signature ($B'>0$) absent from the dry-bulb baseline, where the
night margin is on net contractionary.

\section{Conclusion}
\label{sec:conclusion}

This paper asks which way warming temperatures move agricultural labour, and finds a diurnal
divergence in the results.
To my knowledge, this paper provides the first reduced-form evidence 
and a conceptual framework for understanding the economic channels of diurnal warming --- shifts
in nighttime and daytime temperatures --- on structural transformation. 
The framework carries a single over-identifying restriction: only the daytime margin moves 
local harvest prices. The data confirm this asymmetry ($F=13.34$ testing equality $\btmin=\btmax$ on prices),
providing an empirical grounding for partial market integration, independent from labour
reallocation.
The framework is a guide on the signs rather than magnitudes of the empirical estimates from
a panel and long-differenced approach, which test these comparative statics.
Though the diurnal divergence is robust to a battery of stress tests, 
the empirical approach has a few important nuances and limitations.

First, while multicollinearity of the within-variation ($\rho$) is 
a concern for identification under wet-bulb temperatures $\rho\approx0.84$, in
the surface-air baseline the two margins are within-correlated at only $\rho\approx-0.09$ (online appendix). 
Twoway fixed effects and state trends absorb almost all of that co-movement, leaving the identifying variation 
near-orthogonal (within-transformed $\mathrm{VIF}\approx1$, condition number $\approx1.1$). 
Both $\tmin, \tmax$ carry their sign on agricultural labour shares (+/-, respectively) when specified 
independently with and without rainfall controls in both panel and long differenced estimations.
The within-correlation widens standard errors without drastically altering the magnitude and sign of
the point estimates $\btmin, \btmax$.

Second, the wet-bulb temperature specification is a deterministic nonlinear 
approximation using the raw dry-bulb $\tmin$ and $\tmax$ (online appendix),
mechanically compressing the diurnal temperature range (DTR).
This makes the wet-bulb pair highly collinear relative to the dry-bulb baseline
(within-$\rho\approx0.84$, within-transformed $\mathrm{VIF}\approx3.5$; the online appendix).

Third, specified jointly, $T_{min}$ and $T_{max}$ span the same two-dimensional
regressor space as the daily mean and the diurnal range (DTR), so the diurnal decomposition 
mirrors temperature level vs range effects, uninfluenced by the chosen basis.
A mean-range reparametrisation provides supporting evidence that the DTR carries an independent 
channel on the labour and price margins (online appendix).

Fourth, diurnal warming is a \emph{local} environmental change, compressed or widened by
irrigation, aerosols, and land use. These local changes are arguably endogenous to the 
production function as emphasised by \citet{BarrettOrtizBobeaPham2021}.
Thus, the diurnal margins are not randomly assigned. Irrigation
expansion and other drivers of range compression raise nighttime and lower daytime
temperatures in the same agrarian belts where labour reallocates. Three features in the 
data sample address this confounding: cultivated area contracts identically under both margins,
where an irrigation-driven night signal would instead expand it; the nighttime yield dividend
persists within the historically irrigated belt (online appendix); and measurement drift
in gridded nighttime temperatures near growing settlements would bias the nighttime
coefficient toward zero, rejected by the empirical estimates. Moreover, the asymmetry is rooted in
atmospheric science: the shallow nocturnal boundary layer amplifies the nighttime
response to any change in energy-balance, upstream of local land use
\citep{Davy2016}.

Finally, the conceptual framework maintains labour and land as gross substitutes ($\sigma>1$).
Though it is grounded in the
agricultural-production literature \citep{HayamiRuttan1985, Binswanger1974},
I do not directly estimate this assumption.
The factor-bias is sufficiently derived from non-unit elasticity: under
a Cobb--Douglas production function, the land shock folds into neutral productivity (online appendix).
The consistent signs on labour, area, output and prices discipline this assignment.
Nonlinearity and spatial sorting, held fixed by the linear framework, are left for future research.

A natural extension of the framework's prediction is through India's own colonial-era
and early post-Independence agricultural records, which provide a test for whether a fully closed
subsistence economy without buffers inverts the night-day pattern.
The pattern is consistent in itself: warmer nights hold labour on the farm without the
price and wage distress that warmer days bring, while daytime heat restrains the
non-agricultural transition far more strongly.
Given the two margins carry opposite signs on agricultural labour, the trajectory of
the diurnal temperature range matters as much as the level. 
Where nights are warming faster than days---compressing the range \citep{Karl1991}---the decomposition implies
agricultural consolidation and commercialisation. Where days outpace nights
and the range widens \citep{Zhong2023}, warming implies the double squeeze of farm-labour casualisation and
stalled urban structural transformation.

A back-of-the-envelope calculation---comparing each district's trailing-decade climate at
the 2011 and 1981 Censuses---suggests growing-season nights and days warmed by
$0.45\degree$ and $0.37\degree$C on average over 1981--2011, but the balance is regional---in the districts
where the range compressed, nights ran ahead of days ($+0.71$ versus $+0.17\degree$C), and
where it widened, days led ($+0.54$ versus $+0.21\degree$C). Applying panel estimates
from Table~\ref{Table2}, warming in the range-compressing districts implies a $2.7$-point rise in the
agricultural wage-labour share against an $18\%$ base.
The range-widening $\tmax$ warming implies a slight decline and a $1.0$-point shift onto the casual
seasonal margin. 
Night and day warming, in equivalent magnitudes, sort local agricultural labour in opposite directions.
Under extreme daytime heat, the exit from agricultural labour into informal seasonal employment 
reflects a physiologically constrained churn for survival, rather 
than a sustained pathway to structural change.

\bibliography{\bib}

\clearpage
\appendix
\setcounter{table}{0}
\renewcommand{\thetable}{A\arabic{table}}
\setcounter{figure}{0}
\renewcommand{\thefigure}{A\arabic{figure}}

\section*{Appendix: Summary Statistics and Supporting Exhibits}
\label{app:paper}

This appendix collects the exhibits a reader needs to reproduce the paper's
figures and to check the model's predicted signs against the estimates. The full
battery of robustness checks, the general-equilibrium model with all proofs, and
the inference and measurement diagnostics are in the online appendix.

\begin{table}[!htbp]
    \footnotesize
    \centering
    \caption{Model comparative statics: predicted signs and their empirical counterparts}
    \label{tab:apx_model}
    \renewcommand{\arraystretch}{1.25}
    \begin{tabular}{l cc l}
    \toprule
    & \multicolumn{2}{c}{Predicted sign} & \\
    \cmidrule(lr){2-3}
    Outcome & $T_{min}$ (night) & $T_{max}$ (day) & Empirical counterpart \\
    \midrule
    Cultivated area            & $(-)$        & $(-)$        & Equal contraction; $F=0.01$ (Table~\ref{Table1}) \\
    Harvest price              & $(\approx 0)$ & $(+)$        & Only daytime raises price; $F=13.3$ (Table~\ref{Table1}) \\
    Field wage                 & $(\approx 0)$ & $(-)$        & Daytime wage loss (Table~\ref{Table1}) \\
    Cereal yield               & $(\approx 0,+)$ & $(\approx 0,-)$ & Flat in dry-bulb; $+0.15$ wet-bulb night (online app.) \\
    \addlinespace
    Ag.\ wage-labour share     & $(+)$        & $(-)$        & Night raises, day lowers; $F=24.9$ (Table~\ref{Table2}) \\
    Seasonal share             & $(-)$        & $(+)$        & Opposite of wage labour; $F=14.7$ (Table~\ref{Table2}) \\
    Cultivator share           & $(-)$        & $(\approx 0)$ & Drawn off at night (Table~\ref{Table2}) \\
    \addlinespace
    Rural vs.\ urban amplitude & \multicolumn{2}{c}{same sign, rural $>$ urban} & $F=30.9$ vs.\ $1.9$ (Figure~\ref{fig:ruralurban}) \\
    Towns (non-agriculture)    & $(-)$        & $(-)$        & Both margins depress; $F=2.8$ (Figure~\ref{fig:towns}) \\
    Closed economy             & \multicolumn{2}{c}{signs \emph{reverse}} & Not observed $\Rightarrow$ open ($\chi<\chi^{*}$) \\
    \bottomrule
    \end{tabular}
    \parbox{\linewidth}{
        \vspace{0.5em}
        \footnotesize
        \textit{Notes:}
        Signs are the framework's predictions in the empirically selected partial-open regime
        $0<\chi<\chi^{*}$ under gross substitutes ($\sigma>1$); the price asymmetry is the
        over-identifying sign restriction and the area equality is a calibration rather than a
        theorem. Only signs and the ordering of magnitudes are predicted, not levels. The full
        model, proofs, and the wet-bulb land-augmentation evidence are in the online appendix.
    }
\end{table}

\begin{table}[!htbp]
    \footnotesize
    \centering
    \caption{Summary statistics: district and town, panel and long-difference samples}
    \label{tab:apx_sumstats}
\begin{tabular}{l cc cc}
\toprule
& \multicolumn{2}{c}{Decadal panel (levels)} & \multicolumn{2}{c}{Long difference} \\
& \multicolumn{2}{c}{1981--2011} & \multicolumn{2}{c}{($\Delta$ 2001--11 vs.\ 1981--91)} \\
\cmidrule(lr){2-3}\cmidrule(lr){4-5}
& District & Town & District & Town \\
\midrule
Cultivators (share) & 0.323 & 0.080 & -0.139 & -0.058 \\
 & (0.158) & (0.091) & (0.080) & (0.057) \\
\addlinespace
Agricultural labourers (share) & 0.180 & 0.088 & -0.049 & -0.053 \\
 & (0.111) & (0.097) & (0.057) & (0.060) \\
\addlinespace
Seasonal / marginal (share) & 0.173 & 0.085 & 0.139 & 0.102 \\
 & (0.111) & (0.079) & (0.086) & (0.065) \\
\addlinespace
Non-agricultural (share) & 0.324 & 0.745 & 0.048 & 0.010 \\
 & (0.176) & (0.178) & (0.070) & (0.089) \\
\addlinespace
Nighttime \(T_{min}\) (\degree C) & 17.95 & 18.23 & 0.28 & 0.29 \\
 & (2.39) & (2.22) & (0.24) & (0.24) \\
\addlinespace
Daytime \(T_{max}\) (\degree C) & 29.71 & 29.89 & 0.15 & 0.11 \\
 & (1.95) & (1.63) & (0.19) & (0.20) \\
\addlinespace
Rainfall (growing-season mean) & 1.0 & 1.0 & 0.0 & 0.0 \\
 & (0.5) & (0.5) & (0.1) & (0.1) \\
\addlinespace
\midrule
Units (districts / towns) & 300 & 2,969 & 299 & 2,969 \\
Observations & 1,187 & 11,876 & 299 & 2,969 \\
\bottomrule
\end{tabular}

    \parbox{\linewidth}{
        \vspace{0.5em}
        \footnotesize
        \textit{Notes:}
        Means with standard deviations in parentheses, for the decadal panel (levels,
        1981--2011) and the long-difference cross-section ($\Delta$ 2001--11 vs.\ 1981--91),
        each split into the district (total-population) and Census-town samples. Worker shares
        are of total workers in the unit; temperature and rainfall are growing-season ten-year
        moving averages. The town columns use the balanced panel of towns present in every
        round since 1981---the long-difference sample---so the panel-levels and long-difference
        town statistics describe the same 2{,}969 towns. The district panel counts one more unit
        than the long difference (300 vs.\ 299): one district is observed only at the endline
        and so enters the panel but cannot be differenced.
    }
\end{table}

\begin{table}[!htbp]
    \footnotesize
    \centering
    \caption{District worker reallocation, long difference (1981--2011), by residence}
    \label{tab:apx_distLD}
    \renewcommand{\arraystretch}{1.2}
\begin{tabular}{l *{4}{c} @{\hspace{1.8em}} c}
\toprule
& \multicolumn{4}{c}{Share of Workers} & \\
\cmidrule(lr){2-5}
& \shortstack{Ag. \\ Labour} & \shortstack{Cultivator} & \shortstack{Seasonal} & \shortstack{Non- \\ Agriculture} & \shortstack{Log Total \\ Workers} \\
& (1) & (2) & (3) & (4) & (5) \\
\midrule
\addlinespace
\multicolumn{6}{@{}l}{\textit{Panel A. Total population}} \\
\addlinespace[2pt]
$\Delta T_{min}$ & \textbf{0.075} & -0.024 & \textbf{-0.074} & 0.026 & 0.297 \\
\addlinespace[3pt]
   & (3.58) & (-0.79) & (-2.74) & (0.94) & (1.24) \\
\addlinespace[6pt]
$\Delta T_{max}$ & \textbf{-0.062} & -0.008 & 0.049 & 0.020 & 0.041 \\
\addlinespace[3pt]
   & (-2.51) & (-0.22) & (1.47) & (0.55) & (0.16) \\
\addlinespace[5pt]
\quad F ($\Delta T_{min}=\Delta T_{max}$) & 26.07 & 0.16 & 12.34 & 0.03 & 0.83 \\
\quad \textit{p}-value & $<$0.001 & 0.687 & $<$0.001 & 0.865 & 0.363 \\
\quad Observations & 293 & 293 & 293 & 293 & 293 \\
\midrule
\addlinespace
\multicolumn{6}{@{}l}{\textit{Panel B. Rural population}} \\
\addlinespace[2pt]
$\Delta T_{min}$ & \textbf{0.096} & -0.035 & \textbf{-0.064} & 0.007 & 0.211 \\
\addlinespace[3pt]
   & (4.12) & (-1.11) & (-2.25) & (0.37) & (0.91) \\
\addlinespace[6pt]
$\Delta T_{max}$ & \textbf{-0.068} & 0.011 & 0.056 & -0.001 & 0.025 \\
\addlinespace[3pt]
   & (-2.52) & (0.28) & (1.58) & (-0.02) & (0.10) \\
\addlinespace[5pt]
\quad F ($\Delta T_{min}=\Delta T_{max}$) & 30.87 & 1.24 & 11.36 & 0.07 & 0.48 \\
\quad \textit{p}-value & $<$0.001 & 0.266 & $<$0.001 & 0.794 & 0.489 \\
\quad Observations & 291 & 291 & 291 & 291 & 293 \\
\midrule
\addlinespace
\multicolumn{6}{@{}l}{\textit{Panel C. Urban population}} \\
\addlinespace[2pt]
$\Delta T_{min}$ & \textbf{0.034} & -0.013 & \textbf{-0.028} & 0.007 & 0.632 \\
\addlinespace[3pt]
   & (2.64) & (-0.89) & (-2.09) & (0.25) & (1.75) \\
\addlinespace[6pt]
$\Delta T_{max}$ & 0.006 & 0.015 & 0.006 & -0.028 & 0.305 \\
\addlinespace[3pt]
   & (0.32) & (0.89) & (0.37) & (-0.75) & (0.80) \\
\addlinespace[5pt]
\quad F ($\Delta T_{min}=\Delta T_{max}$) & 1.89 & 1.69 & 3.23 & 0.69 & 0.60 \\
\quad \textit{p}-value & 0.171 & 0.194 & 0.073 & 0.407 & 0.439 \\
\quad Observations & 291 & 291 & 291 & 291 & 293 \\
\midrule
Rainfall Control & Y & Y & Y & Y & Y \\
State FE & Y & Y & Y & Y & Y \\
\bottomrule
\end{tabular}

    \parbox{\linewidth}{
        \vspace{0.5em}
        \footnotesize
        \textit{Notes:}
        Long-difference estimates of nighttime ($\Delta T_{min}$) and daytime
        ($\Delta T_{max}$) growing-season surface-air warming on district worker shares,
        decomposed into the total, rural, and urban populations of each district (the
        within-district residence split). These are the coefficients behind
        Figure~\ref{fig:ruralurban}: the agricultural wage-labour column of Panels~B and~C is
        plotted there. Point estimates with $|t|>2$ in bold over district-clustered
        $t$-statistics; the $F$-statistic and its $p$-value test
        $\Delta T_{min}=\Delta T_{max}$. All specifications include state fixed effects and a
        precipitation control.
    }
\end{table}

\begin{table}[!htbp]
    \footnotesize
    \centering
    \caption{Town worker reallocation, long difference (1981--2011)}
    \label{tab:apx_townLD}
    \renewcommand{\arraystretch}{1.2}
\begin{tabular}{l *{4}{c} @{\hspace{1.8em}} c}
\toprule
& \multicolumn{4}{c}{Share of Workers} & \\
\cmidrule(lr){2-5}
& \shortstack{Ag. \\ Labour} & \shortstack{Cultivator} & \shortstack{Seasonal} & \shortstack{Non- \\ Agriculture} & \shortstack{Log Total \\ Workers} \\
& (1) & (2) & (3) & (4) & (5) \\
\midrule
\addlinespace
$\Delta T_{min}$ & \textbf{0.063} & -0.009 & -0.016 & \textbf{-0.038} & 0.272 \\
\addlinespace[4pt]
   & (5.46) & (-0.86) & (-1.35) & (-2.16) & (1.61) \\
\addlinespace[8pt]
$\Delta T_{max}$ & \textbf{0.041} & 0.022 & 0.017 & \textbf{-0.078} & 0.241 \\
\addlinespace[4pt]
   & (2.59) & (1.49) & (1.15) & (-3.47) & (1.41) \\
\addlinespace
Observations & 2,969 & 2,969 & 2,969 & 2,969 & 2,969 \\
F-stat ($\Delta T_{min}=\Delta T_{max}$) & 1.61 & 3.38 & 4.06 & 2.78 & 0.11 \\
\quad \textit{p}-value & 0.205 & 0.067 & 0.044 & 0.096 & 0.738 \\
\midrule
Rainfall Control & Y & Y & Y & Y & Y \\
FE / Trend & Y & Y & Y & Y & Y \\
\bottomrule
\end{tabular}

    \parbox{\linewidth}{
        \vspace{0.5em}
        \footnotesize
        \textit{Notes:}
        Census-town worker reallocation under nighttime ($\Delta T_{min}$) and daytime
        ($\Delta T_{max}$) growing-season surface-air warming, across the balanced panel of
        Census towns---the coefficients behind Figure~\ref{fig:towns}. The four mutually
        exclusive worker categories (agricultural wage labour, cultivators, seasonal,
        non-agricultural) sum to one, so the share coefficients sum to approximately zero;
        column~5, set apart, reports the long-run change in log total town workers. Point
        estimates with $|t|>2$ in bold over district-clustered $t$-statistics; the
        $F$-statistic and its $p$-value test $\Delta T_{min}=\Delta T_{max}$. State fixed
        effects and a precipitation control throughout.
    }
\end{table}

\end{document}